# Robust autofocusing propagation in turbulence

Nana Liu[1], Liu Tan[1], Kaijian Chen[1], Peilong Hong[2,3,6], Xiaoming Mo[1], Bingsuo Zou[4], Yi Liang[1,5,7]

[1]*Guangxi Key Lab for Relativistic Astrophysics, Center on Nanoenergy Research, School of Physical Science and Technology, Guangxi University, Nanning, Guangxi 530004, China*
[2]*School of Mathematics and Physics, Anqing Normal University, Anqing, Anhui 246133, China*
[3]*The MOE Key Laboratory of Weak-Light Nonlinear Photonics, Nankai University, Tianjin 300457, China*
[4]*School of Physical Science and Technology and School of Resources, Environment and Materials, Key Laboratory of new Processing Technology for Nonferrous Metals and Materials, Guangxi University, Nanning 530004, China*
[5]*State Key Laboratory of Featured Metal Materials and Lifecycle Safety for Composite Structures, Nanning 530004, China*
[6]*plhong@uestc.edu.cn*
[7]*liangyi@gxu.edu.cn*

**We conducted a comprehensive study on the robust propagation of the same spot-size Gaussian beam (SSGB), same envelope Gaussian beam (SEGB), Circular Airy beam (CAB) and Circular Pearcey beam (CPB) in complex environments. Our findings clearly demonstrate that autofocusing beams exhibit higher stability in propagation compared with the Gaussian beams. To validate our results, we statistically analyze the intensity fluctuation of autofocusing beams and Gaussian beams. The analysis reveals that the intensity fluctuations of autofocusing beams are significantly smaller than that of Gaussian beams. Furthermore, we study the variation of the coherence factor and find that the coherence of autofocusing beams is better than that of Gaussian beams under turbulence. Additionally, we observe the change of the scintillation index (SI) with the propagation distance *z*, and our results show that autofocusing beams exhibit less oscillation than Gaussian beams, indicating that autofocusing beams propagate in complex environments with less distortion and fewer intensity fluctuation. Overall, our results suggest that autofocusing beams are promising for applications such as optical trapping and manipulation in complex environments. Our study provides valuable insights into the selection of stable beams that exhibit high intensity and high field gradient at the focal position in complex environments.**

## 1. Introduction

Due to the strong absorption and scattering, complex environments such as atmospheric turbulence[1, 2] and complex biological media[3] have always been a major obstacle to the beams propagation in optical communication[4], optical trapping and manipulation[3, 5]. To overcome this challenge, various methods have been developed. For example, the use of structured light including Bessel-Gaussian, Hermite-Gaussian, Laguerre-Gaussian, vortex beams and vector beams, have been shown to be effective in mitigating the effects of turbulence on optical systems[1, 2]. While most current researches on structured light in turbulence have been focused on optical communication, structured light is also promising in optical trapping and manipulation [6]. This is especially true when optical tweezer works in complex environments, where the scattering and turbulence pose challenges to the optical trapping and manipulation[3].

As we know, excellent focusing of a beam is critical for good optical trapping and manipulation in complex environments. At present, the most popular method is to reshape the wavefront of beams and focus beams through the random medium [7-10]. However, to recover the optimized wavefront requires careful pre-calibration on the scattering process. Thus, another choice is to apply structured light directly. In recent years, the autofocusing beams[11, 12] have gained popularity in optical tweezers and manipulation [3, 6, 13-18]. Without traditional lenses or nonlinear effects, the autofocusing beams allow sudden autofocusing, giving rise to high intensity at the focus. Previous researches have demonstrated that Gaussian beams are susceptible to severe distortions and continuous "center of mass" hopping under turbulent conditions, while Airy beams experience significantly less distortion and drift [19, 20]. Moreover, when a Gaussian beam passes through a scattering medium, random speckles form due to light scattering [21], whereas Airy beams maintain their basic characteristics, except for a reduction in intensity due to scattering. These findings suggest that the autofocusing beams originating from the Airy beams such as Circular Airy beam (CABs), may have greater robustness during propagation in complex environments.

Compared to CAB, Circular Pearcey beam (CPB), which is derived from another kind of accelerating beam called Pearcey beam[22-24], offers better autofocusing performance. CPB has a shorter autofocusing distance and stronger peak focusing intensity, and it eliminates post-focus oscillations, making it more effective for optical manipulation. Moreover, circular Pearcey vortex beam in complex environments can perform better than circular Airy vortex beam theoretically[25], implying CPB may also present a similar better result than CAB.

Therefore, to confirm the potentials of CAB and CPB in optical manipulation under complex environments, we theoretically and experimentally investigate how the autofocusing propagation of CAB and CPB changes with the degree of air turbulence. Moreover, we also compare CAB, CPB with equivalent Gaussian beams selected according to two different criteria. One is the same focal spot-size Gaussian beam (SSGB), the other is the same envelope Gaussian beam (SEGB), whose beam width equals to the diameters of main rings of CAB and CPB[17, 26]. We analyze the distortion and intensity fluctuation of these beams passing through complex environments and compare the robustness in propagation between these beams. Our results show that the autofocusing beams can resist some effects of air turbulence and light scattering, implying a potential robustness in penetrating tissues or media for excellent optical trapping and manipulation. Furthermore, our findings indicate that CPB performs better than SSGB, SEGB and CAB in turbulence. Overall, our study highlights the potential of autofocusing beams as a superior option for applications that require robustness in complex environments.

## 2. Theoretical background

The input fields of GB, CAB and CPB are expressed as [23, 27]:

$$\psi_{GB}(r, 0) = A_0 \times \exp[-\frac{r^2}{w_0^2}], \quad (1)$$

$$\psi_{CAB}(r, 0) = A_1 \times Ai[\frac{c_1-r}{w_1}] \times \exp[\alpha(\frac{c_1-r}{w_1})] \times q(r), \quad (2)$$

$$\psi_{CPB}(r, 0) = A_2 \times Pe[\frac{c_2-r}{w_2}, 0] \times q(r), \quad (3)$$

where $A_0$, $A_1$, $A_2$ are the amplitudes. $Ai(x) = 1/\pi \int_0^\infty exp[i(1/3t^3 + xt)] dt$ is the Airy function, $Pe(u, v) = \int_{-\infty}^{\infty} exp[i(t^4 + t^2u + tv)] dt$ is the Pearcey function, $r = \sqrt{x^2 + y^2}$ is the radial coordinate, $w_0$ is the waist radius of Gaussian beam, $c_1, c_2$ are parameters controlling the initial radius of CAB and CPB, $w_1, w_2$ are the transverse scale factors, $\alpha$ is the exponential decay factor, $q(r) = \text{rect}(\frac{r}{2r_1})(r \geq 0)$ is an indicator used to limit the power and $r_1$ is used to limit the beam region. For comparing the propagation behavior effectively and consistently, we adopted appropriate parameters to make beams have same focal lengths and spot sizes or initial envelops (See Supplementary Section 1 for detail). In addition, $\lambda = 532\,nm$, input power $P = 1W$.

In experiment, SSGB and SEGB are focused by an additional lens (L3 in the 1st row of Fig. 1(a), focal length equals to the autofocusing length $f_z$=300 mm), while CAB and CPB are generated by the off-axis hologram method like Ref. [14], as depicted in the 2nd row Fig. 1(a). We used an expanded quasi-plane Gaussian beam to transmit through a spatial light modulator (SLM), and loaded the desired hologram of the CAB/CPB on the SLM (hologram is manifested by computing the interference between the CAB/CPB wave and a plane wave), then a 4-$f$ system (composed by L3 and L4) with a spatial filter (only let the 1st order of light modulated by the SLM pass by) turned the beam into CAB/CPB. The propagations of the generated SSGB, SEGB, CAB and CPB were monitored by an image lens and a charge-coupled device (CCD). To simulate atmospheric turbulence, we heated an accordion-shaped aluminum foil with an alcohol lamp to generate turbulent air currents in the beam's path[19]. We controlled the turbulence by varying the lamp position to adjust the hotplate's temperature.

In theory, to analyze comprehensively the propagation properties in turbulence, we employed the widely accepted Kolmogorov turbulence theory[28] to calculate the beam propagation for comparing the experimental results. In this case, the refractive index structure parameter $C_n^2$ is usually employed to define the strength of atmospheric turbulence[29]. Then, based on multiple-phase-screen method[30, 31], we can simulate the beam propagation in atmospheric turbulence. The modified von *Kármán* power spectral density (PSD) in Kolmogorov turbulence theory is expressed as[2, 29]:

$$\phi_n^{mvk}(\kappa) = 0.033 C_n^2 \frac{\exp(-\frac{\kappa^2}{\kappa_m^2})}{(\kappa^2+\kappa_0^2)^{11/6}} \quad (4)$$

where $\kappa = 2\pi(f_x\hat{i} + f_y\hat{j})$ is angular spatial frequency in rad/m. The larger $\kappa$ is, the smaller the PSD $\phi_n^{mvk}(\kappa)$ is, and the influence of turbulence is smaller. $\kappa_m = 5.92/l_0$, $\kappa_0 = 2\pi/L_0$, $l_0$ is called the inner scale, $L_0$ is called the outer scale. Finally, under the paraxial condition of Fresnel diffraction, we simulated the beam propagation via the Angular Spectrum Method.

## 3. Results and discussion

### 3.1 Beams Propagation in turbulence

The theoretical simulation results are presented in Figs. 1(b-e), showing that SSGB, SEGB, CAB and CPB experience different intensity fluctuations when passing through the same distance with the same turbulence. The sideview results of the theoretical simulation in Figs. 1(b1-e1) reveal that, compared to the case of no turbulence (See Figs. S1 in the Supplement), the peak intensities of SSGB, SEGB and CAB in turbulence of $C_n^2 = 10^{-8}$ are attenuated by 50%, 43%, 41%, while that of CPB decreases only 30%. At the focal plane, the instantaneous peak intensity of SSGB and SEGB deviate farther from the original center position than CAB and CPB [Figs. 1(b2-e2) and the Visualizations 1-4]. These observations suggest that autofocusing beams exhibit a more robust propagation in turbulence than Gaussian beams, while they maintain autofocusing. Also, we conducted experiments to further demonstrate such behavior of these beams in atmospheric turbulence. The experimental profile results match the theoretical results well, as shown in the insets of Figs. 1(b1-e1) and 1(b2-e2), as well as supplementary Visualizations 5-8.

To further support our findings, we recorded the instantaneous dynamical positions of the beams in simulation and experiment, illustrating their statistical distributions in Figs. 1(b3-e3) and 1(b4-e4). Seemingly, autofocusing beam's distributions are more concentrated than Gaussian beams, implying a smaller centroid deviation. Actually, our calculations reveal that the corresponding statistical positions variances of SSGB, SEGB, CAB and CPB are 0.00361, 0.00407, 0.00344,

0.00333 in simulation, and 0.00138, 0.00222 0.00126, 0.00122 in experiment. This clearly indicates that CPB exhibits a most robust focusing propagation. A similar trend persists even under stronger turbulence conditions such as $C_n^2 = 10^{-7}$. For example, the longitudinal focus position of SSGB and SEGB exhibits significant shifts (from 300mm to 240, 180mm) while that of CAB and CPB remains relatively stable with minor changes (from 300mm to 264, 270 mm).

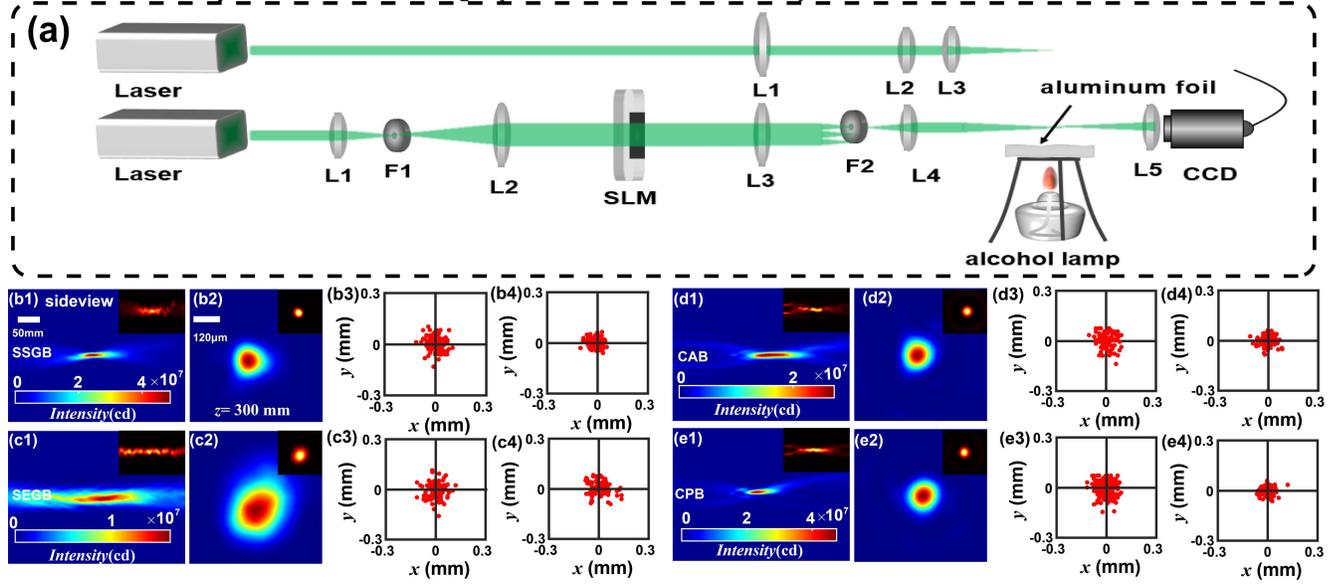

Fig. 1. (a) Schematic diagram of the experiment setup. L: Lens; F: Filter; SLM: Spatial Light Modulator; CCD: Charge-coupled Device. (b-e) Simulated (Jet) and experimental results (Hot) of propagation of SSGB, SEGB, CAB and CPB in atmospheric turbulence. (b1-e1) Sideview propagation, (b2-e2) intensity distributions at the focus $z = f_z$ ($f_z$ is the focal length). (b3-e3) and (b4-e4) the statistical distributions of instantaneous dynamical positions of the beams in simulation and experiment.

### 3.2. The coherence factor

Light scattering in complex environments causes beam distortion and intensity fluctuation, which leads to a loss of coherence. To assess the impact of turbulence on coherence, we also calculate the modulus of the complex coherence factor (MCF) of the beam[29, 32]( The detail see Section 3 in the Supplement).

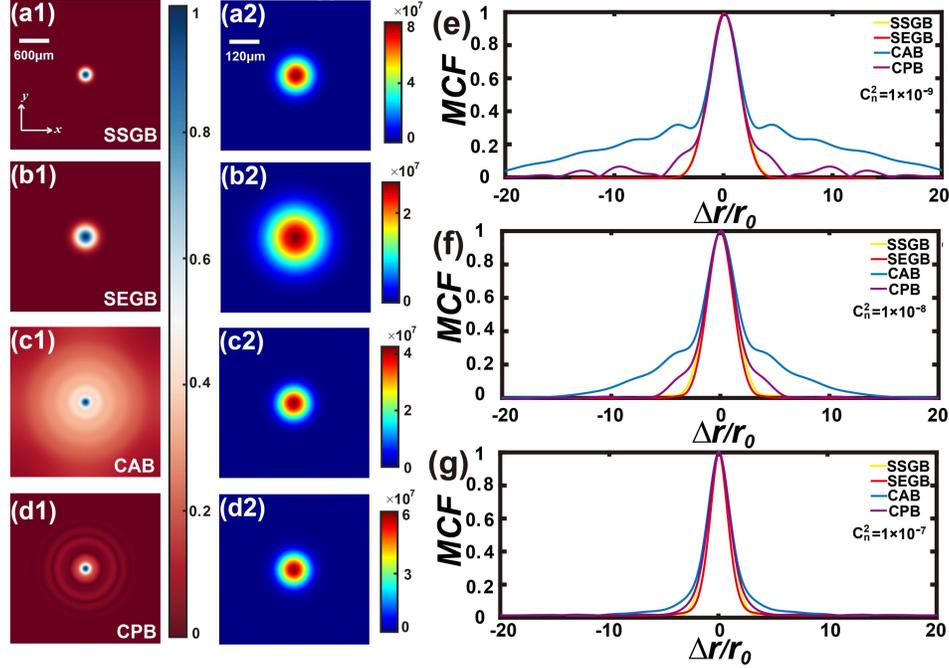

Fig. 2. MCF and average intensity profiles of SSGB, SEGB, CAB and CPB. (a-d) two-dimensional cases of $C_n^2 = 10^{-9}$. (e-g) are the variations of MCF in one-dimensional form with the ratio of the axial change $\Delta r$ to the spot radius $r_0$ in the case of $C_n^2 = 10^{-9}$, $10^{-8}$ and $10^{-7}$.

Figures 2(a-d) depict the variation of the MCFs and average intensity distributions for SSGB, SEGB, CAB and CPB at the focus, considering $C_n^2$ values of $10^{-9}$, $10^{-8}$ and $10^{-7}$. It's evident that, when contrasted with turbulence-free conditions, all beams demonstrate an amplified average intensity distribution under the influence of turbulence, as depicted in Figs. 2(a2-d2). Furthermore, a comparative examination of the MCFs reveals that the distributions for CAB and CPB are more extensive compared to those of Gaussian beams [Figs. 2(a1-d1)]. Since the coherence factor of the beam is also impacted by beam spreading during short-distance propagation [33], we plotted the change of MCF with respect to $\Delta r/r_0$ ($\Delta r = r - r'$, and $r_0$ is the focus radius) for accessing the more accurate perturbation of turbulence on beams' coherence, as shown in Figs. 2(e-g). One can see that, the MCFs of autofocusing beams consistently outpace those of Gaussian beams, even though all MCFs exhibit rapid decline with escalating turbulence [Figs. 2(e-g)]. In other words, autofocusing beams exhibit a more robust propagation.

### 3.3. The scintillation index

Scintillation is the intensity fluctuation caused by random perturbation [34] when beam passes through turbulent, which is harmful to the beam propagation. Usually, scintillation index (SI) is employed to describe such scintillation, which can be calculated by [2, 30]:

$$\sigma(r,z)^2 = \frac{\langle I(r,z)^2 \rangle}{\langle I(r,z) \rangle^2} - 1 \qquad (5)$$

where $\sigma^2$ represents the variance of irradiance, $I$ is the instantaneous intensity, and the angle brackets stand for the average value.

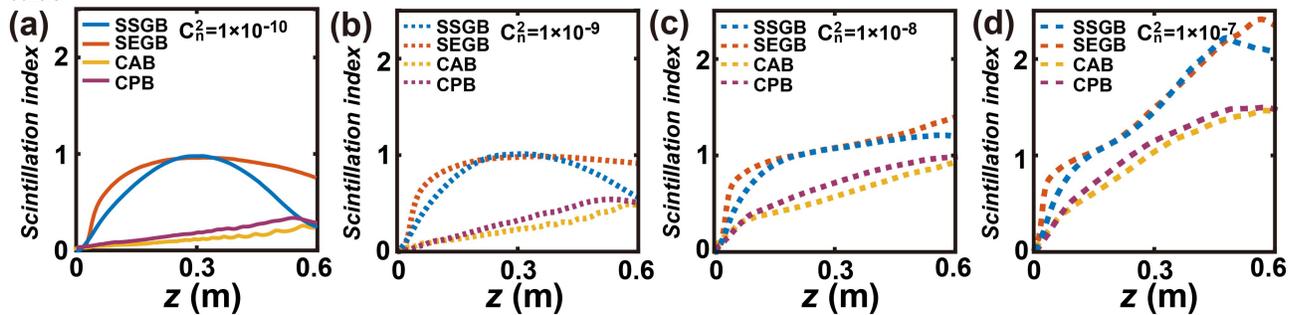

**Fig. 3.** Scintillation indexes of SSGB, SEGB, CAB and CPB vary with the propagation distance $z$ under different index structure parameter $C_n^2$: (a) $10^{-10}$, (b) $10^{-9}$, (c) $10^{-8}$, (d) $10^{-7}$.

Figures 3(a-d) portray the average intensity fluctuations (SI) of the entire beams encompassing SSGB, SEGB, CAB and CPB, considering varying structure parameters $C_n^2$. As expected, there's a noticeable rise in SI as the index structure parameter increases, along with an increase in propagation distance. In this case, SI in stronger turbulence rises quickly even at lower distances. Notably, for Gaussian beams, the SI demonstrates a peak at a specific propagation distance, followed by a subsequent decline. This pattern arises due to the auto compensation effects or the diffractive spreading of fragments within the distorted beams, as mentioned in Ref. [25, 35]. Anyway, across all turbulence scenarios, the SI values for SSGB and SEGB consistently surpass those of CAB and CPB in all turbulence cases. This discrepancy highlights the heightened sensitivity of Gaussian beams to intensity fluctuations when subjected to the impact of turbulence. This finding reinforces our previous assertion, once again showing that the propagation of autofocusing beams remains relatively more stable in the presence of atmospheric turbulence.

### 3.4. Discussion

The results presented above demonstrate that both CAB and CPB exhibit robust focusing propagation than Gaussian beams in turbulent environments. This can be explained as follows:

Airy beams and Pearcey beams exhibit a remarkable property known as self-healing within the realm of structured light [19, 23]. This intrinsic ability empowers them to rapidly restore their optical field intensity distribution, effectively reverting to a state closely resembling the configuration prior to encountering any obstruction. In more detail, based on catastrophe theory, we know that the phenomenon of the acceleration of Airy beams and Pearcey beams originates from the presence of their catastrophe caustics, i.e., fold and cusp caustics. Caustics are regions where the intensity reaches its highest level, and they are closely associated with the existence of stable "singularities"[36, 37]. Actually, such "singularities" here are positions where the second derivative of the full phase of the beam angular spectrum is zero, identified as caustic. Moreover, the structural stability is a characteristic feature inherent to caustics of Airy and Pearcey beams [38]. In this case, the beams achieve equilibrium, with their inherent characteristics remaining stable against minor disturbances, including subtle phase shifts induced by turbulence. Consequently, these beams exhibit unique attributes such as self-healing properties and a heightened ability to propagate robustly, particularly in turbulent environments. Thus, CAB and CPB, stemming from Airy beams and Pearcey beams, surpass Gaussian beams in terms of their propagation capabilities within turbulent environments.

Moreover, as presented in Figs. 2(e-g) and 3 in contrast to CPB, as turbulence intensifies (shifting from $C_n^2 = 10^{-9}$ to $C_n^2 = 10^{-7}$), CAB exhibits wider MCF width and smaller SI, but changes more rapidly. This points to CAB's heightened susceptibility to strong turbulence impact. So, for robust turbulence scenarios, CPB emerges as the superior option for autofocusing propagation due to its stronger autofocusing peak intensity and more stable focus position.

Our result can be extended to other complex environments. For example, when beams pass through a scattering medium of polystyrene solution, autofocusing beams still show less distortion than Gaussian beam (See Section 5 in the Supplement).

## 4. Conclusion

In summary, we have conducted a comprehensive investigation on the propagation behavior of SSGB, SEGB, CAB and CPB in complex environments. Our results indicate that autofocusing beams stand out with superior robust autofocusing performance—showing less intensity distortion and fluctuation compared to Gaussian beams. Statistical analysis at the focal position supports autofocusing beams' heightened stability, particularly CPB. The coherence factor analysis confirms that autofocusing beams exhibit greater coherence than Gaussian beams, reinforcing their superior turbulence resistance. Examining scintillation variation with propagation distance, we find that autofocusing beams experience fewer disruptions and smaller scintillation than Gaussian beams, underscoring their enhanced intensity stability. These findings hold practical significance, guiding the selection of beams that remain stable, maintain high focal intensity, and effectively propagate in challenging environments. This study finds relevance in diverse applications like optical communication, imaging and optical trapping.

**Funding**

Supported by National Natural Science Foundation of China (11604058), Guangxi Natural Science Foundation (2020GXNSFAA297041, 2020GXNSFDA238004), Innovation Project of Guangxi Graduate Education (YCSW2023083), Sichuan Science and Technology Program(2023NSFSC0460) and Open Project Funding of the Ministry of Education Key Laboratory of Weak-Light Nonlinear Photonics(OS22-1).

**Disclosures**

The authors declare no conflicts of interest.

**Data availability**

Other data underlying the results presented in this paper are not publicly available at this time but may be obtained from the authors upon reasonable request.

**Supplemental document**

See Supplement for supporting content including selection parameters of beams, beams propagation without turbulence, the calculation of coherence factor, the scintillation index of beams and beams propagation in the scattering medium.

# Supplemental document

## 1. Selection parameters of SSGB, SEGB, CAB and CPB

In order to ensure a meaningful and consistent comparison of propagation behaviors, we carefully selected suitable parameters to align the beams' focal lengths and spot sizes, or their initial envelopes. Specifically, we maintained the radius of the main rings and the focal length of both CAB and CPB at identical values, as outlined below:

Indeed, it's well-known that the Airy function Ai(x) and Pearcey function Pe(x,0) attain their maximum peak positions at x= -1.018 and x=-2.199, respectively. Therefore, it is straightforward to deduce that the parameters associated with the same main rings of CAB and CPB meet the following criteria:

$$c_1 - r_B = -1.018 w_1 \quad (S1)$$
$$c_2 - r_B = -2.199 w_2 \quad (S2)$$

Here, $c_1$, $c_2$ are two parameters controlling the initial radiuses of main rings of the autofocusing beams, $r_B$ is the radius of the main ring, $w_1$ $w_2$ are the transverse scale factors.

Furthermore, the radius $r_B$ of the main ring of CAB and CPB, which can be represented in terms of their respective autofocusing length $f_z$. This relationship is deduced from the trajectory expressions found in Refs. [S1] for Airy beams and [S2] for Pearcey beams. Consequently, if the autofocusing lengths for both CAB and CPB are kept consistent, we arrive at the following relationship:

$$r_B = \frac{f_z^2 \lambda^2}{16\pi^2 w_1^3} \quad (S3)$$
$$r_B = \left(\frac{f_z \lambda}{6\pi}\right)^{3/2} \frac{4}{w_2^2} \quad (S4)$$

Eqs. (S3) and (S4) are for CAB and CPB, respectively. Then, based on Eqs. (S1-S4), we can harmonize the properties of CAB and CPB to ensure they possess identical main ring radii and autofocusing lengths.

In our study, we have defined $c_2 = 0$ and $w_2 = 114.5$μm. Thus, we set the focal length of CPB as $f_z = 300$mm and radius of main ring as $r_B = 255$μm. To achieve parity with these parameters of CPB including $f_z$ and $r_B$, we fine-tuned the parameters of CAB, adjusting $c_1 = 170$μm, $w_1 = 93$μm as determined by equations (S1) and (S3). Finally, CAB and CPB have the same ring radii and autofocusing lengths.

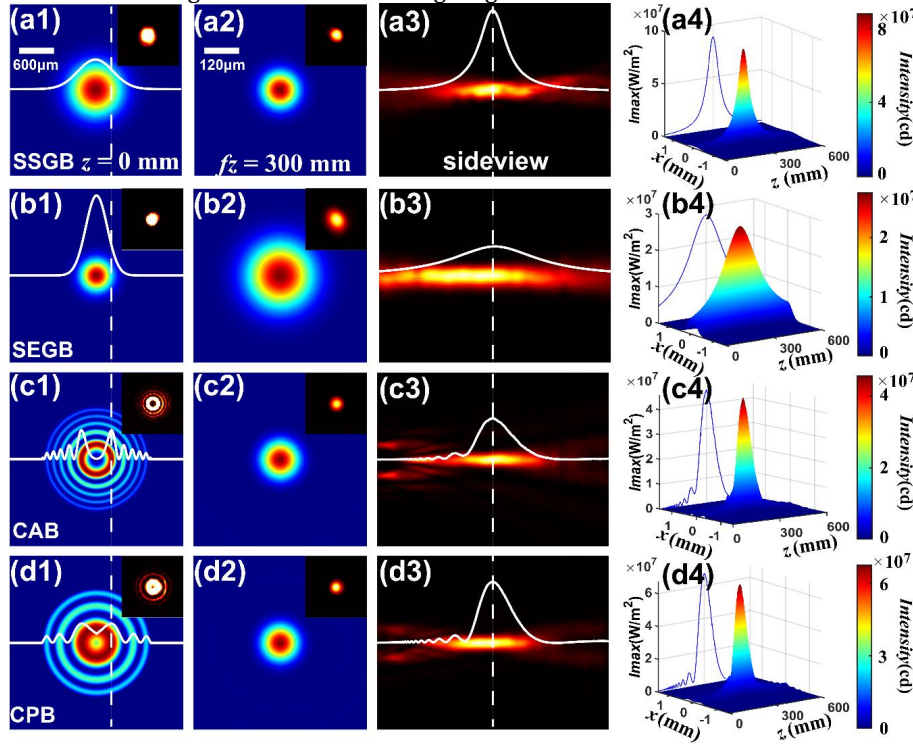

Fig. S1 Simulated (Jet colorbar) and experimental results (Hot colorbar) without turbulence. The intensity distribution of the SSGB (a1-a4), SEGB (b1-b4), CAB (c1-c4) and CPB (d1-d4). (a1-d1) at the initial plane z = 0 mm, (a2-d2) at the focal plane $z = f_z$, (a3, a4-d3, d4) the sideview of propagation, showing the different propagation properties of these beams. The white curves in the middle are the corresponding simulated intensity distribution curves. The white line in (a1-d1) represents the same envelope Gaussian beam (SEGB), whose beam width of SEGB is equal to the diameters of main rings. The white line in (a3-d3) represents the focus position of these beams.

Next, we generate equivalent Gaussian beams chosen using two distinct criteria. As elucidated in the paper, one criterion is based on the selection of beams with the same focal spot-size (SSGB), while the other involves beams with the same envelope (SEGB), whose beam width matches the main ring diameters. For SSGB, we adjusted the waist radius $w_0 = 425 \; \mu m$ to achieve a Gaussian beam with a matching focal spot size after being directed through a lens with a focal length $f_z = 300 mm$. For SEGB, we set the Gaussian beam's waist radius $w_0 = r_B$ to align with the main ring radius, and then propagated it through a lens with a focal length $f_z = 300$ mm.

Consequently, we have successfully generated beams with equivalent focal lengths and spot sizes, or initial envelopes, as visually presented in Figure S1.

## 2. Beams propagation without turbulence

When $C_n^2 = 0$, signifying an absence of turbulence, the theoretical simulation outcomes are illustrated in Figures S1(a-d). Notably, SSGB and autofocusing beams exhibit identical focal spot sizes [Figures S1(a2) and S1(c2-d2)], and the beam width for SEGB corresponds to the diameters of the main rings of autofocusing beams [Figures S1(b1-d1)]. Upon propagating to a distance of z=300 mm, SSGB, SEGB, CAB and CPB converge into a single point marked by a peak intensity [Figures S1(a2-d2) and S1(a3-d3)], i.e., these beams have a same focal length. Subsequent to this focus, the peak intensities of these beams rapidly diminish [Figures S1(a4-d4)]. These outcomes closely align with experimental findings, as depicted in the insets of Figure S1. Through meticulous theoretical calculations, it has been determined that the intensity ratio between the peak intensity ($8.80 \times 10^7 \; W/m^2$) of SSGB and the maximum intensity ($1.72 \times 10^6 \; W/m^2$) at the initial plane is approximately 51 times. Conversely, the intensity ratios for SEGB, CAB and CPB are approximately 6 times, 27 times and 60 times, respectively. Notably, the CPB demonstrates the highest maximum intensity ratio, implying that it might exhibits superior focusing and propagation characteristics compared to SSGB, SEGB and CAB in turbulence-free conditions.

## 3. The calculation of coherence factor

After passing the turbulent air, the optical field $\psi(r)$ of the beam is obtained by Angular Spectrum Method. The description of the mutual coherence function can be expressed as[S3]:

$$\Gamma(r,r^{'},z) = \langle \psi(r)\psi^{*}(r^{'}) \rangle \quad (S5)$$

where $r$ and $r^{'}$ are the coordinates of two spatial points, from the mutual coherence function we can compute the modulus of the complex coherence factor (MCF) as

$$\mu(r,r^{'},z) = \frac{|\Gamma(r,r^{'},z)|}{[\Gamma(r,r,z)\Gamma(r^{'},r^{'},z)]^{1/2}} \quad (S6)$$

## 4. The scintillation index of beams

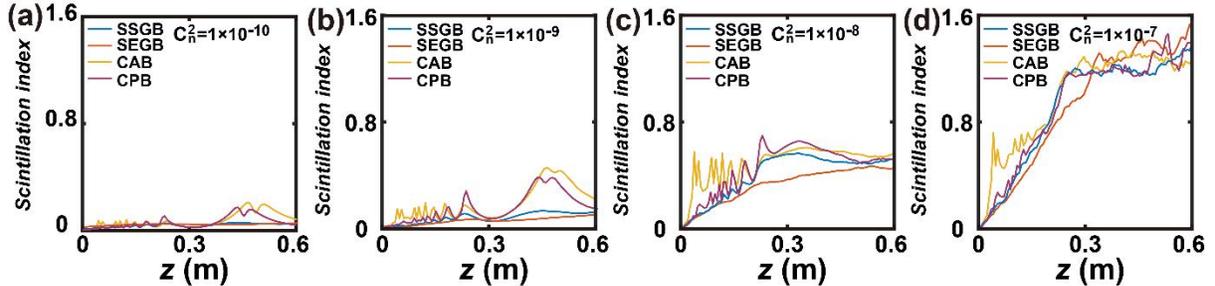

Fig. S2 Scintillation indexes of the central position SSGB, SEGB, CAB and CPB vary with the propagation distance $z$ under different index structure parameter $C_n^2$: (a) $C_n^2 = 10^{-10}$, (b) $C_n^2 = 10^{-9}$, (c) $C_n^2 = 10^{-8}$, (d) $C_n^2 = 10^{-7}$.

The fluctuations in intensity (SI) at the central position of SSGB, SEGB, CAB and CPB are depicted in Figures S2(a-d) for varying structural parameter $C_n^2$ values. As anticipated, the SI increases with higher $C_n^2$ values and as the propagation distance grows. Notably, we observed that while the SI of SSGB and SEGB is initially smaller than that of CAB and CPB in weak turbulence, it exhibits more significant changes with increasing turbulence. This suggests that Gaussian beams experience substantial intensity fluctuations under the influence of strong turbulence. In the majority of turbulence scenarios, CPB consistently exhibits a smaller SI compared to CAB.

Moreover, during propagation—especially under weak turbulence conditions—the presence of multiple focal points contributes to oscillations in the scintillation index (SI). This phenomenon becomes evident in the side view profile illustrated in Figures S1(c3-d3), where the emergence of multiple focal points aligns with several peaks observed in the scintillation index at the center position of the beams. However, given the intricate nature of turbulence, it's crucial to acknowledge that maintaining a consistent alignment between individual peaks of the scintillation index and the side view propagation may not persist. Remarkably, under stronger $C_n^2$ values, these focal points undergo slight shifts of a few millimeters, inducing minor variations in the focusing positions. The heightened SI observed at the focal point originates from the concentration of light perturbed by the entirety of the turbulence. Our speculation suggests that turbulence effects may obscure the peaks of the scintillation index, particularly when subjected to conditions of strong turbulence, as visually demonstrated in Fig. S2. Indeed, similar phenomena have also been observed in other studies, as mentioned in Refs. [25,35] of the manuscript. These effects could potentially arise from auto compensation processes or the diffractive spreading of fragmented beams within the context of turbulence.

## 5. Beams propagation in the scattering medium

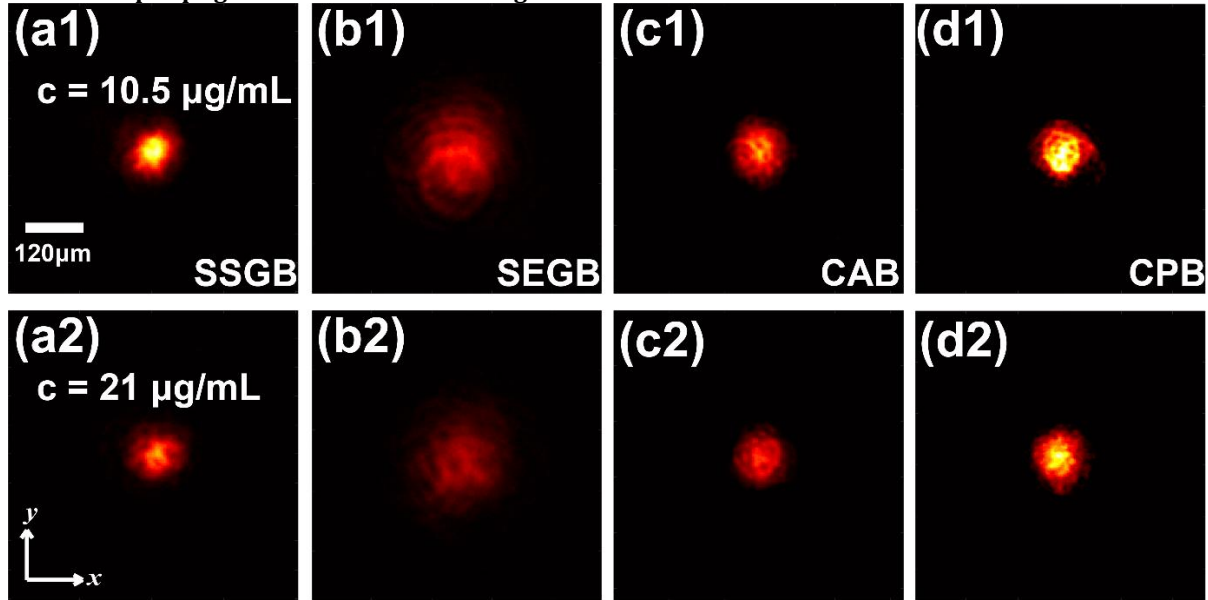

Fig. S3 SSGB (a1-a2), SEGB (b1-b2), CAB (c1-c2) and CPB (d1-d2) through the scattering medium. (a1-d1) through a low concentration polystyrene solution, (a2-d2) through a high concentration polystyrene solution.

The behavior of light propagation is inherently influenced by the disorderly arrangement of particles within a random medium, often resulting in light scattering and the deformation of light beams. Nonetheless, achieving a stable and consistent propagation of light beams over extended distances is of paramount importance. In our pursuit to delve into the distortion characteristics of SSGB, SEGB, CAB and CPB, we employed polystyrene particles at varying concentrations as the scattering medium[S4]. To accomplish this, we suspended two distinct concentrations of polystyrene particles within pure water. The solution underwent ultrasonic treatment to ensure the even dispersion of particles, subsequently being positioned within a cupola setup—an alternative to the alcohol lamp utilized in our prior experiments. Notably, the particle diameter measured 2.5μm, chiefly leading to Mie scattering effects[S4]. The concentrations of the two polystyrene solutions were set at 10.5 μg/mL and 21 μg/mL. Then, through this setup, we could observe the behavior of these beams as they traversed through the particle suspensions.

Figures S3(a1-d1) and S3(a2-d2) illustrate the propagation behavior of SSGB, SEGB, CAB and CPB, respectively, within polystyrene solutions of varying concentrations, as observed at the focal position. As the concentration of the polystyrene solution increases, the intensity of these beams diminishes while the degree of distortion becomes more pronounced. Importantly, it's noteworthy that Gaussian beams exhibit a more pronounced scattering effect than autofocusing beams within polystyrene solutions, regardless of the solution's concentration. These outcomes collectively suggest that autofocusing beams, especially CPBs, showcase heightened stability within scattering mediums.

It's crucial to note that in this scattering experiment, we solely controlled the scattering of the beam's focus within the polystyrene solution. The beam propagated through free space before entering the polystyrene solution, thereby allowing us to manage the propagation distance within complex environments. Based on these observations, we can confidently conclude that autofocusing beams demonstrate superior stability within scattering mediums. This conclusion holds value as it underscores the enhanced performance of autofocusing beams under conditions involving beam distortion and scattering.

## 6. Animation and videos:

Supplementary videos show beams propagation of SSGB, SEGB, CAB and CPB in turbulence.